\documentclass[prc,twocolumn,superscriptaddress,showpacs,twoside,floatfix]
{revtex4-1}

\usepackage{amssymb,epsfig}

\begin{document}

\title{Democratic Decay of $^{6}$Be Exposed by Correlations}

\author{I.A.~Egorova}
\affiliation{Bogolyubov Laboratory of Theoretical Physics, JINR, Dubna, 141980
Russia}
\author{R.J.~Charity}
\affiliation{Departments of Chemistry and Physics, Washington University,
St.~Louis, Missouri 63130, USA}
\author{L.V.~Grigorenko}
\affiliation{Flerov Laboratory of Nuclear Reactions, JINR, Dubna, RU-141980
Russia}
\affiliation{GSI Helmholtzzentrum f\"{u}r Schwerionenforschung, Planckstra{\ss}e
1, D-64291 Darmstadt, Germany}
\affiliation{National Research Center ``Kurchatov Institute'', Kurchatov Square 
1, RU-123182 Moscow, Russia}
\author{Z.~Chajecki}
\affiliation{National Superconducting Cyclotron Laboratory and
Department of Physics and Astronomy, Michigan State University,
East Lansing, Michigan 48824, USA}
\author{D.~Coupland}
\affiliation{National Superconducting Cyclotron Laboratory and
Department of Physics and Astronomy, Michigan State University,
East Lansing, Michigan 48824, USA}
\author{J.M.~Elson}
\affiliation{Departments of Chemistry and Physics, Washington University,
St.~Louis, Missouri 63130, USA.}
\author{T.K.~Ghosh}
\affiliation{Variable Energy Cyclotron Centre, 1/AF Bidhannagar,
Kolkata 700064, India}
\author{M.E.~Howard}
\affiliation{Department of Physics and Astronomy, Rutgers University,
New Brunswick, New Jersey 08903, USA}
\author{H.~Iwasaki}
\affiliation{National Superconducting Cyclotron Laboratory and
Department of Physics and Astronomy, Michigan State University,
East Lansing, Michigan 48824, USA}
\author{M.~Kilburn}
\affiliation{National Superconducting Cyclotron Laboratory and
Department of Physics and Astronomy, Michigan State University,
East Lansing, Michigan 48824, USA}
\author{Jenny Lee}
\affiliation{National Superconducting Cyclotron Laboratory and
Department of Physics and Astronomy, Michigan State University,
East Lansing, Michigan 48824, USA}
\author{W.G.~Lynch}
\affiliation{National Superconducting Cyclotron Laboratory and
Department of Physics and Astronomy, Michigan State University,
East Lansing, Michigan 48824, USA}
\author{J.~Manfredi}
\affiliation{Departments of Chemistry and Physics, Washington University,
St.~Louis, Missouri 63130, USA}
\author{S.T.~Marley}
\affiliation{Department of Physics, Western Michigan University, Kalamazoo,
Michigan 49008, USA}
\author{A.~Sanetullaev}
\affiliation{National Superconducting Cyclotron Laboratory and
Department of Physics and Astronomy, Michigan State University,
East Lansing, Michigan 48824, USA}
\author{R.~Shane}
\affiliation{Departments of Chemistry and Physics, Washington University,
St.~Louis, Missouri 63130, USA}
\author{D.V.~Shetty}
\affiliation{Department of Physics, Western Michigan University, Kalamazoo,
Michigan 49008, USA}
\author{L.G.~Sobotka}
\affiliation{Departments of Chemistry and Physics, Washington University,
St.~Louis, Missouri 63130, USA}
\author{M.B.~Tsang}
\affiliation{National Superconducting Cyclotron Laboratory and
Department of Physics and Astronomy, Michigan State University,
East Lansing, Michigan 48824, USA}
\author{J.~Winkelbauer}
\affiliation{National Superconducting Cyclotron Laboratory and
Department of Physics and Astronomy, Michigan State University,
East Lansing, Michigan 48824, USA}
\author{A.H.~Wuosmaa}
\affiliation{Department of Physics, Western Michigan University, Kalamazoo,
Michigan 49008, USA}
\author{M.~Youngs}
\affiliation{National Superconducting Cyclotron Laboratory and
Department of Physics and Astronomy, Michigan State University,
East Lansing, Michigan 48824, USA}
\author{M.V.~Zhukov}
\affiliation{Fundamental Physics, Chalmers University of Technology, S-41296
G\"{o}teborg, Sweden}


\begin{abstract}
The interaction of an $E/A$=70-MeV $^7$Be beam with a Be target was used to
populate levels in $^6$Be following neutron knockout reactions. The three-body
decay of the ground and first excited states into the $\alpha$+$p$+$p$ exit
channel were detected in the High Resolution Array. Precise three-body
correlations
extracted from the experimental data allowed us to obtain insight
into the mechanism of the three-body democratic decay. The correlation data are
in a good agreement with a three-cluster-model calculation and thus validate
this theoretical approach over a broad energy range.
\end{abstract}


\maketitle

\begin{figure}
\includegraphics[width=0.48\textwidth]{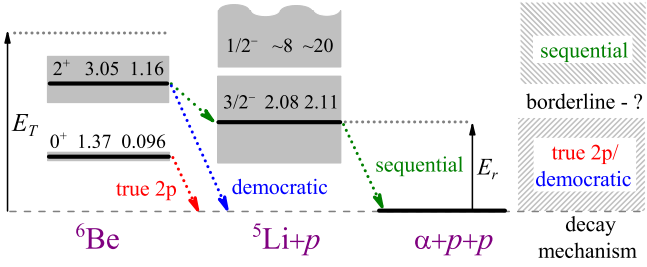}
\caption{Level and decay scheme for $^{6}$Be and illustrations of possible
decay mechanisms. The continuum states are labeled $\{J^{\pi},E_r,\Gamma \}$ }
\label{fig:scheme}
\end{figure}

\textit{Introduction.}
%
%
--- The $^{6}$Be system is located beyond the proton dripline, and its ground and 
excited states all belong to the three-body $\alpha$+$p$+$p$ continuum. 
Moreover, the $^{6}$Be ground state might be considered a so-called ``true 
two-proton emitter''; a system for which one-proton decay is energetically 
prohibited and thus should emit two protons simultaneously, as most of 
the strength for $^5$Li intermediate states is inaccessible (Fig.\ 
\ref{fig:scheme}). However, at large excitation energies, one expects that the 
decay mechanism in the three-body continuum should eventually evolve to a 
sequential decay process through such intermediate states. In  light two-proton 
emitters, these intermediate states are often quite broad and hence the concept 
of ``democratic decay'' was proposed \cite{Bochkarev:1989,Pfutzner:2012}. 
``Democracy'' in this case means that no strong focusing in kinematical space is 
produced even if the intermediate states are accessible for decay; the decay 
mechanism remains essentially three-body in nature. The three-body decay of the 
$^6$Be ground state may thus be classified as both a ``true'' and a 
``democratic'' two-proton decay. The interplay and transition between the 
different decay mechanisms in three-body systems have been strongly debated and 
they are still not completely understood 
\cite{Pfutzner:2012,Danilin:1987,Barker:2003,Blank:2008,Alvarez:2008,%
Grigorenko:2009,Grigorenko:2009c}. The location of the borderline between the 
three-body decay dynamics (true $2p$ or democratic) and two-body dynamics 
(sequential decay) is not known.

In recent years there has been a revival of interest in the  $^{6}$Be system
\cite{Grigorenko:2009,Grigorenko:2009c,Papka:2010,Charity:2010,Fomichev:2012}
with comparative studies to two-proton radioactive decay in $^{45}$Fe
\cite{Grigorenko:2009}, precise studies of correlations for the ground state
\cite{Grigorenko:2009c}, and the discovery of an ``isovector soft dipole mode''
in a charge-exchange reaction \cite{Fomichev:2012}. $^{6}$Be is the lightest
two-proton ground-state emitter and, being relatively easily accessible in
experiments, could become a benchmark system for studies of  two-proton
emission (two-proton radioactivity in heavier nuclei). In addition, because of
isopin symmetry, the two-proton correlations can shed light on the structure of
the mirror neutron-halo nucleus $^{6}$He \cite{Grigorenko:2009c}.

In this Letter, we report on studies of the $^{6}$Be continuous spectrum up to a
decay energy of $E_T \sim 10$ MeV (the $E_T$ is energy above the $\alpha$+$p$+$p$
threshold). The high-statistical-significance and high-resolution data  provide a very
detailed view of the evolution of the correlation patterns with excitation
energy. This allows us to obtain  insights into the mechanism of two-proton
decay. The result is a demonstration of the counterintuitive character of the
evolution of the decay mechanism with excitation energy.

\textit{Experiment.}
%
%
--- A primary beam of $E/A$=150 MeV $^{16}$O was extracted from the Coupled
Cyclotron Facility at the National Superconducting Cyclotron Laboratory at
Michigan State University with an intensity of 125 pnA. This beam bombarded a
$^{9}$Be target, and $^7$Be projectile-fragmentation products were selected by
the A1900 separator with a momentum acceptance of $\pm 0.5\%$. This  $^7$Be
secondary beam had an intensity of 4$\times $10$^{7}$~s$^{-1}$ with a purity of
$\sim90\%$. It impinged on a 1-mm thick target of $^9$Be, creating $^6$Be
projectile-like fragments via neutron knockout reactions.

The protons and $\alpha$ particles created following $^6$Be decay were detected
in the High Resolution Array \cite{Wallace:2007}. For this experiment, the array consisted
of 14 $\Delta E$-$E$ [Si-CsI(Tl)] telescopes located at a distance 90~cm
downstream from the target and subtended zenith angles from 1.4$^\circ$ to
13$^\circ$. Each telescope consisted of a 1.5-mm thick, double-sided Si strip
$\Delta E$ detector followed by a 4-cm thick CsI(Tl) $E$ detector. The $\Delta
E$ detectors are 6.4$\times $6.4~cm in area with each of the faces divided
into 32 strips. Each $E$ detector consisted of four separate CsI(Tl) elements,
each spanning a quadrant of the preceding Si detector. Signals produced in the
896 Si strips were processed with the HINP16C chip electronics
\cite{Engel:2007}.

The energy calibration of the Si detectors was obtained with a $^{228}$Th
$\alpha$-particle source. The particle-dependent energy calibrations of the
CsI(Tl) detectors were achieved with $E/A$ = 60 and 80 MeV beams of protons and
$\alpha$ particles selected with the A1900 separator. An experimental $^6$Be
decay energy $E_T$ was determined from the invariant mass of each detected
$\alpha$+2\textit{p} event minus the rest masses of the three decay products.


\textit{Theoretical model.}
%
%
--- The dynamics of the three-body $\alpha$+$p$+$p$ continuum of $^{6}$Be is
described by solving  the inhomogeneous three-body Schr\"odingier equation
for wave functions (WF) with the outgoing asymptotic
\begin{equation}
(\hat{H}_3 - E_T)\Psi^{(+)} = \Phi_{\mathbf{q}},
\label{eq:shred}
\end{equation}
corresponding to an approximate boundary condition of the three-body Coulomb
problem. The differential cross section is expressed via the flux induced by the
WF $\Psi^{(+)}$ on the remote surface $S$ by
\begin{equation}
\frac{d \sigma}{d^3k_{\alpha}d^3k_{p_1}d^3k_{p_2}} \sim  \left. \langle
\Psi^{(+)} | \hat{j} | \Psi^{(+)} \rangle \right|_S.
\label{eq:cress-sect}
\end{equation}
To compare to experimental data, the calculated sevenfold-differential cross
sections were used in Monte Carlo (MC) simulations of the experiment, taking
into account the apparatus bias and resolution. The model is described in detail
in Ref. \cite{Grigorenko:2009c} and applied in different ways in Refs. \cite{Grigorenko:2009,Fomichev:2012}.

\begin{figure}
\includegraphics[width=0.48\textwidth]{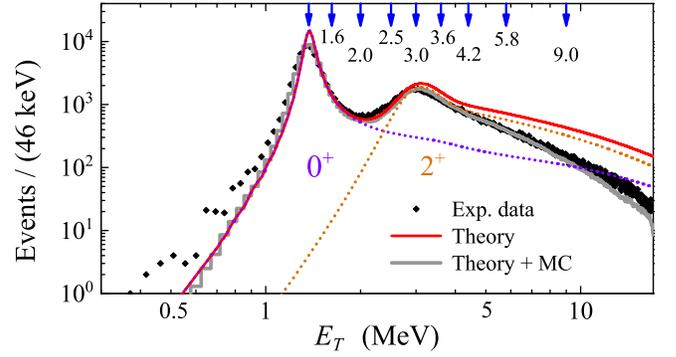}
\caption{Experimental invariant-mass spectrum of $^{6}$Be and the
fitted theoretical result. Within this work, the data points show experimental
data, the solid curve shows the theoretical result, and the gray histogram shows the Monte
Carlo simulation that takes into account the bias introduced by the 
experimental setup. To the extent that Monte Carlo simulation reproduces the
experiment result, the theoretical curve can be interpreted as the
reconstructed distribution.
The dotted  curves
show the contributions of $0^+$ and $2^+$ states to the theoretical spectrum.
The arrows indicate the boundaries of the energy bins used in this work.}
\label{fig:inv-mass}
\end{figure}

The source function $\Phi_{\mathbf{q}}$ for the reaction considered in this work
is constructed in the approximation of a sudden removal of a neutron from
$^{7}$Be (transparent limit of the Serber model),
\begin{equation}
\Phi_{\mathbf{q}} = \int d^3 r_n e^{i\mathbf{q r}_n} \langle
\Psi_{^4\text{\scriptsize
He}} | \Psi_{^7\text{\scriptsize Be}} \rangle ,
\label{eq:sour}
\end{equation}
where $\mathbf{r}_n$ is the radius vector of the removed neutron and
$\mathbf{q}$ is the transferred momentum.
 The $^{7}$Be WF
is constructed in the spirit
of cluster orbital shell model approximation (e.g., Ref. \cite{Zhukov:1994}) as an ``inert''
$\alpha$ core plus
a neutron and two protons occupying $p_{3/2}$ and $p_{1/2}$ configurations with
coupling $[l_j(\nu)[l_j(\pi_1)l_j(\pi_2)]_J]_{J_{7\text{Be}}}$,
\begin{eqnarray}
\Psi_{^7\text{\scriptsize Be}} & = & \Psi_{^4\text{\scriptsize
He}} \bigl( \alpha
[p_{3/2} [p^2_{3/2}]_0]_{3/2} + \beta [p_{3/2} [p^2_{1/2}]_0]_{3/2} \nonumber \\
& + & \gamma [p_{3/2} [p^2_{3/2}]_2]_{3/2} + \delta [p_{3/2}
[p_{3/2}p_{1/2}]_2]_{3/2} \bigr).
\label{eq:psi-7be}
\end{eqnarray}
Neutron removal  populates the $0^+$ state in $^{6}$Be for terms with
coefficients $\{\alpha, \beta\}$ and populates the $2^+$ state for terms with
coefficients $\{\gamma, \delta\}$. The ratios $\alpha/\beta$ and $\gamma/\delta$
control the spin contents of the source terms in Eq.\ (\ref{eq:shred}).
From the fit of the experimental $E_T$ distribution in Fig.\ \ref{fig:inv-mass}
we obtain $\{ \alpha,\beta,\gamma,\delta \}=\{0.42,0.3,0.49,0.7\}$.
The sensitivity of the reaction to the structure of
$^{7}$Be is an interesting question by itself that will be discussed elsewhere.

\begin{figure*}
\centerline{
\includegraphics[width=0.48\textwidth]{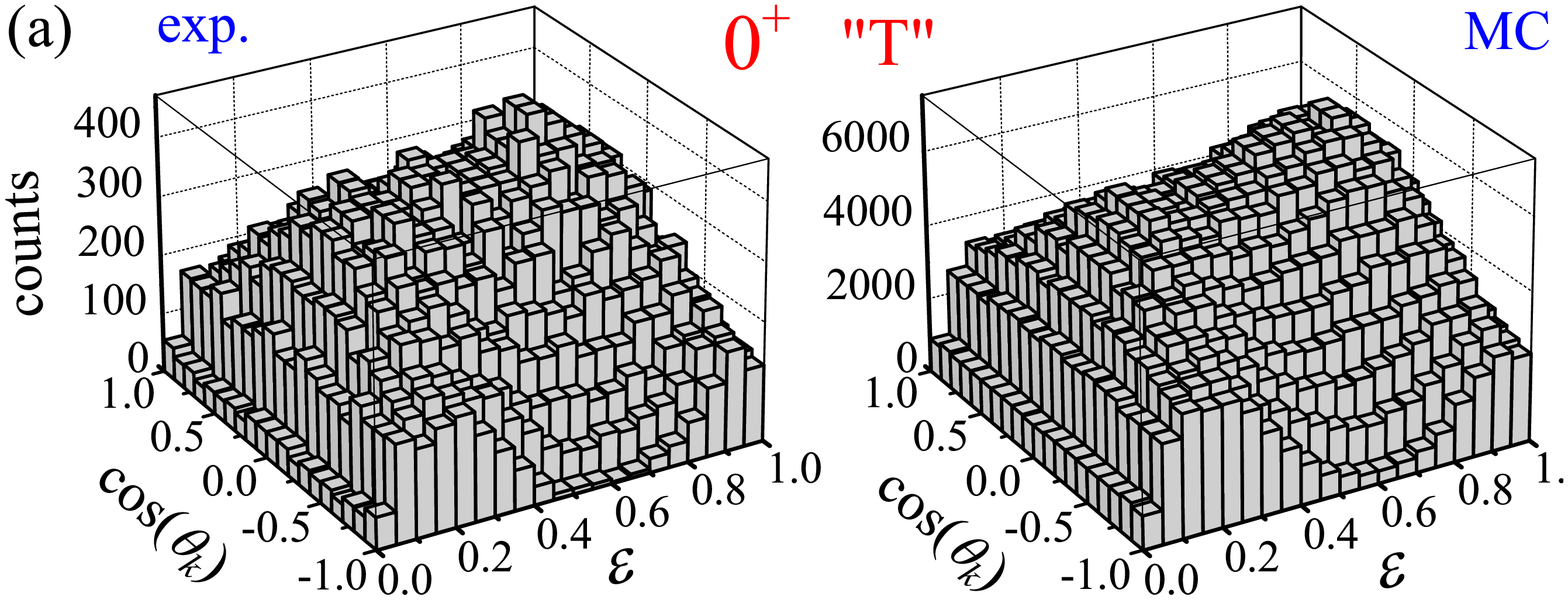} $\qquad$
\includegraphics[width=0.48\textwidth]{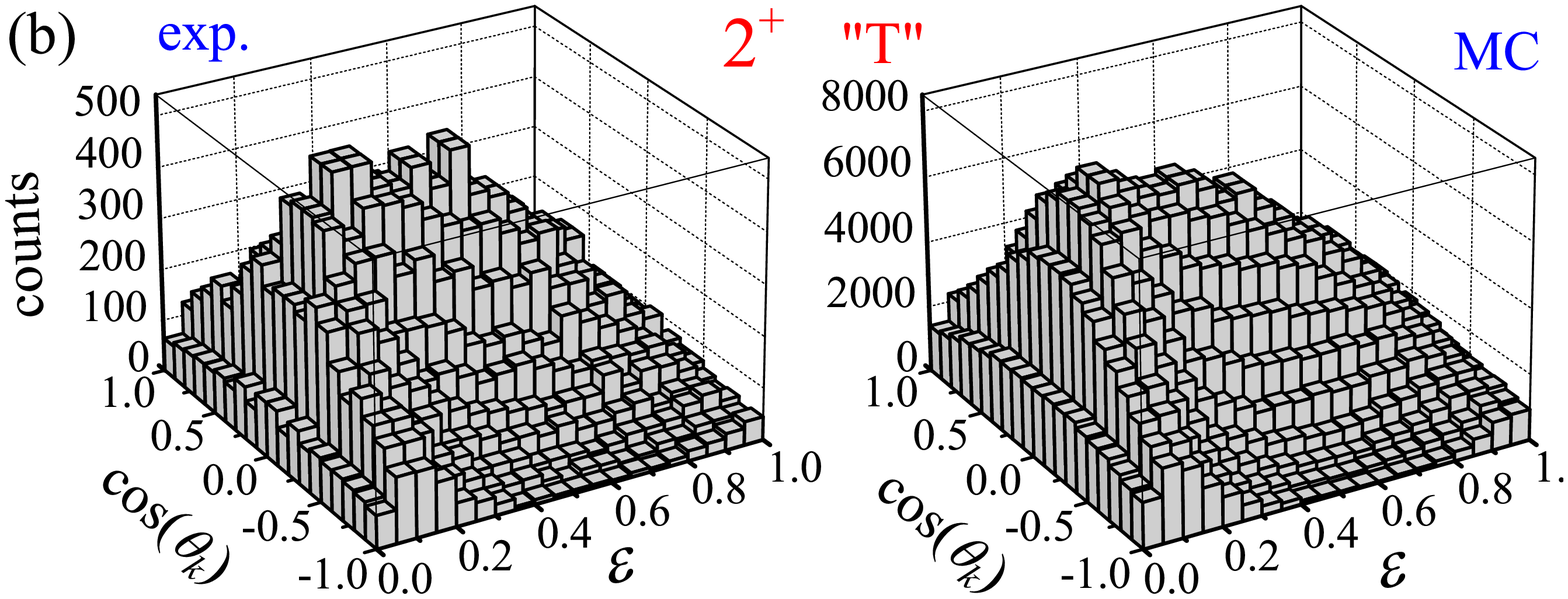}
}
\smallskip
\centerline{
\includegraphics[width=0.48\textwidth]{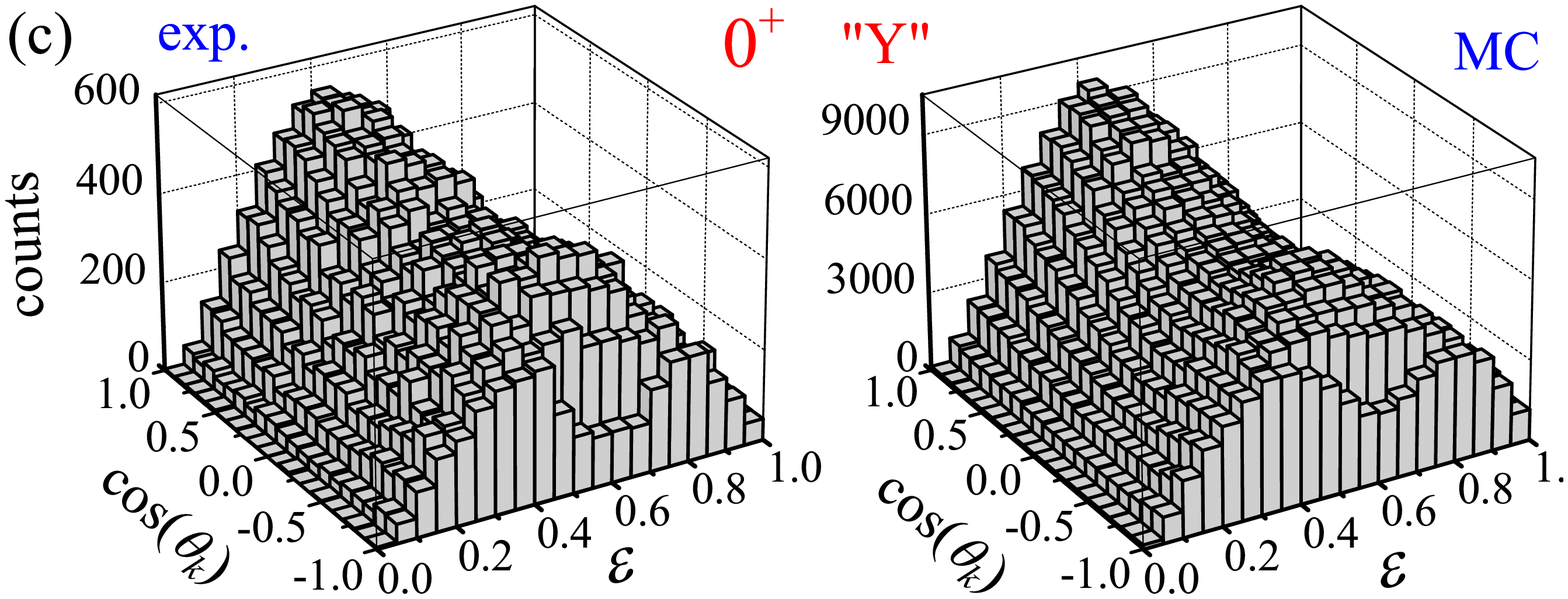} $\qquad$
\includegraphics[width=0.48\textwidth]{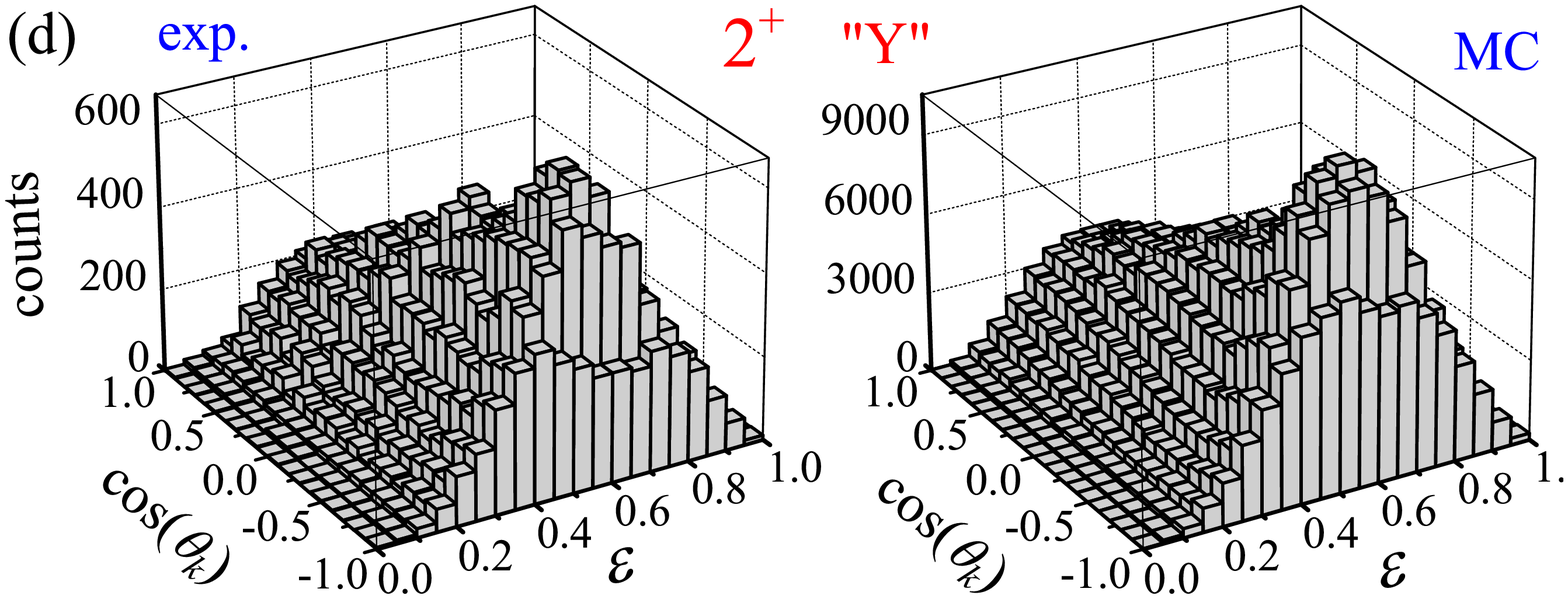}
}
\caption{Complete energy-angular correlations for the the $0^+$ [(a) and (c)] and
$2^+$ [(b) and (d)] states of $^{6}$Be. Comparison of experiment and MC simulations for
Jacobi ``T'' and ``Y'' systems (upper and lower rows, respectively). The data
are taken for 0.5 and 1 MeV wide bins centered on the $0^+$ and $2^+$ resonance
peaks, respectively.}
\label{fig:lego}
\end{figure*}


\textit{Complete energy-angular correlations.}
%
%
--- Two-body decays are described by just two quantities, energy and width.
For three-body decays, one also needs at least two extra continuous degrees of
freedom that in this work are the  energy distribution parameter
$\varepsilon $ and the angle $\theta_{k}$ between the Jacobi momenta
$\mathbf{k}_{x}$, $\mathbf{k}_{y}$,
\begin{eqnarray}
\varepsilon = E_x/E_T \, ,\qquad \cos(\theta_k)=(\mathbf{k}_{x} \cdot
\mathbf{k}_{y}) /(k_x\,k_y) \, , \nonumber \\
{\bf k}_x  =  \frac{A_2 {\bf k}_1-A_1 {\bf k}_2 }{A_1+A_2} \, ,  \,\;
{\bf k}_y  =  \frac{A_3 ({\bf k}_1+{\bf k}_2)-(A_1+A_2) {\bf k}_3}
{A_1+A_2+A_3} , \nonumber \\
E_T =E_x+E_y=k^2_x/2M_x + k^2_y/2M_y ,  \qquad
\label{eq:corel-param}
\end{eqnarray}
where $M_x$ and $M_y$ are the reduced masses of the $X$ and $Y$ subsystems (see 
e.g., \ Ref.\ \cite{Grigorenko:2009c} for details). If we put $k_3 \rightarrow
k_{\alpha}$, then the correlations are obtained in the ``T'' Jacobi system where
$\varepsilon$ describes the energy correlation in the $p$-$p$ channel. However,
if we put $k_3 \rightarrow k_{p}$, then the correlations are obtained in one of
the ``Y'' Jacobi systems where $\varepsilon$ describes the energy correlation in
the $\alpha$-$p$ channel.

The calculated energy-angular distributions are in excellent agreement with the
experimental data for both the $0^+$ and $2^+$ resonances; see  Fig.\ \ref{fig:lego}. 
These correlation data for the two resonance states are also in
agreement with the recent results from Refs.\ 
\cite{Grigorenko:2009,Grigorenko:2009c,Charity:2010,Fomichev:2012} and  with the
older data  of Refs.\ \cite{Geesaman:1977,Bochkarev:1989}. Only the data of 
Ref.\ \cite{Papka:2010} were found to be inconsistent. However, compared to all these
other data sets, the present data have the highest statistical significance and
thus provide the best validation of the theoretical model. In addition to the
present high-statistical-significance data, we are able to explore the evolution of the
correlations on and off resonance.

\begin{figure}
\includegraphics[width=0.49\textwidth]{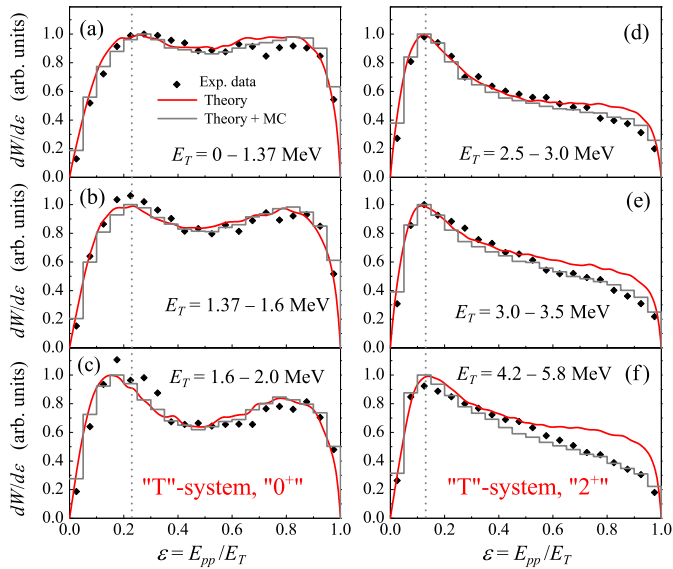}
\caption{Evolution of energy distribution in the Jacobi ``T''
system (between two protons) with the decay energy. The left and right columns
show the energy ranges where the $0^+$ and $2^+$ states dominate. Vertical
dotted lines are shown to help one evaluate the shift or lack of shift
in the peak location between different panels.
See the caption of Fig.\ \ref{fig:inv-mass}
for explanation of symbols and curves.}
\label{fig:t-evol}
\end{figure}

\textit{Evolution of energy distribution between two protons.}
%
%
--- Figure \ref{fig:t-evol} shows the evolution of the distribution of relative
energy between two protons  with $E_T$. There is a qualitative
difference between the distributions for the $0^+$ [Figs.~\ref{fig:t-evol}
(a) and \ref{fig:t-evol} (b)] and $2^+$ 
[Figs.~\ref{fig:t-evol} (d) and \ref{fig:t-evol} (e)] states. In addition, the
small-$E_{pp}$ region for $p$-$p$ motion becomes enhanced with increasing $E_T$
for the $0^+$ state. This result is unexpected as the $p$-$p$ final-state
interaction (FSI) is generally considered to be a predominantly low-energy
phenomenon, but this trend is also confirmed in the calculations.
For the first time, we can see the evolution of the distributions in the
transition-energy region [Fig.\ \ref{fig:t-evol} (c)] characterized by strong
$0^+/2^+$ state mixing. For the energy region covering the $2^+$ state  and
beyond [Fig.\ \ref{fig:t-evol} (d)--\ref{fig:t-evol} (f)], the energy distributions demonstrate
stable shapes far beyond the 2$^+$ peak [Fig.\ \ref{fig:t-evol} (f)], a result
again confirmed in the calculations.


\textit{Evolution of energy distribution between alpha and proton.}
%
%
--- There is a widespread belief that as soon as the intermediate state becomes
energetically accessible, the decay mechanism changes over from three-body decay
to a sequential decay through this resonance. To see what happens in reality,
let us consider the energy correlation in the $\alpha$-$p$ channel, which should
reflect the $^{5}$Li ground-state resonance in the case of sequential decay.

\begin{figure}
\includegraphics[width=0.49\textwidth]{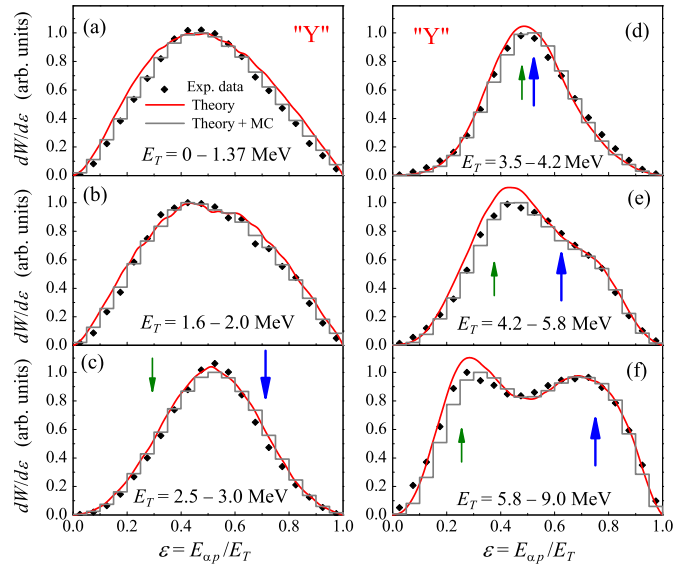}
\caption{Evolution of the energy distribution in the Jacobi ``Y''
system (relative energy between the alpha and one of the protons) with energy $E_T$.
Arrows indicate the positions of the $^{5}$Li ground-state resonance in the three-body
energy window. See the caption of Fig.\ \ref{fig:inv-mass} for explanation of symbols
and curves.}
\label{fig:y-evol}
\end{figure}

We can see in Figs.\ \ref{fig:y-evol} (a) and \ref{fig:y-evol} (b) that at low $E_T$, the shapes of the
energy distribution in the Jacobi ``Y'' system have a  relatively broad
bell-like profile typical for true $2p$ decay \cite{Pfutzner:2012}. However as
$E_T$ increases the profile first becomes  significantly narrower. This
narrowing happens exactly when the $^{5}$Li ground-state resonance enters the
decay window; Fig.\ \ref{fig:y-evol} (c). The location of sequential-decay
strength to the centroid of the $^{5}$Li resonance, $E_{\alpha p}=E_r$($^5$Li)
and $E_{\alpha p} \approx E_T-E_r$($^5$Li), where the concentration of strength
might intuitively be expected, is indicated in Fig.\ \ref{fig:y-evol} by large
blue and small green arrows, respectively. It seems that for $E_T < 2
E_r$($^5$Li) , the availability of the two-body $\alpha$-$p$ resonance for
sequential decay does not lead to  correlation patterns that one might consider
typical of sequential decay with two peaks or a peak plus a shoulder.
Significant evidence for such sequential correlations are only observed when
$E_T \gtrsim 2 E_r(^5$Li$) +\Gamma (^5$Li).

\begin{figure}[b]
\includegraphics[width=0.46\textwidth]{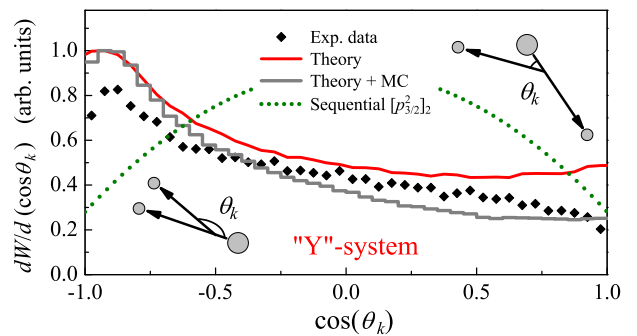}
\caption{Angular distribution in the Jacobi ``Y'' system for the
$E_T=5.8-9$ MeV
bin.}
\label{fig:y-cost-high}
\end{figure}

Let us now turn to the energy correlation at high $E_T$ values ($5.8 <E_T< 9.0$
MeV), Fig.\ \ref{fig:y-evol} (f). The $^{5}$Li energy correlation is very
evident here with peaks located at  the energies indicated by the two arrows.
However, if sequential decay is the only process here, then the \emph{angular}
correlations should be completely defined by angular-momentum coupling. The
predicted angular distribution corresponding to sequential decay via $[p_{3/2}
\otimes p_{3/2}]_2$ coupling (dotted curve) is compared to the corresponding
experimental data in Fig.\ \ref{fig:y-cost-high}. In contrast to this
prediction, the experimental distribution has a strong asymmetry with a focusing
of the two protons at small relative angles. Technically, such asymmetry cannot
exist for pure sequential decay and must be connected with an interference
between odd/even parity configurations  (say of $[p^2]$ with $[sd]$
configurations in $^{6}$Be). Physically, it is clear that the peak at
$\cos(\theta_k) \sim -1$ is connected with $p$-$p$ FSI present in a realistic
Hamiltonian [Eq.\ (\ref{eq:shred})].

\begin{figure}
\includegraphics[width=0.48\textwidth]{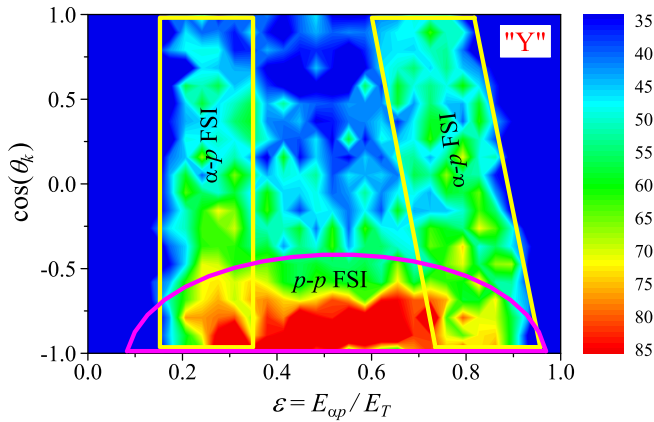}
\caption{Experimental joint energy-angular distributions in the
Jacobi ``Y'' system for the $E_T=5.8-9$ MeV bin. The legend shows the number of
events in each element the $30 \times 20$ grid.}
\label{fig:y-eps-cost-high}
\end{figure}

A more complete picture of the decay is obtained by studying the joint
energy-angular distribution of Fig.\ \ref{fig:y-eps-cost-high}. This
distribution contains regions clearly identifiable with $p$-$p$ and $\alpha$-$p$
FSIs and, in addition, a broad transition region. Each of these regions is
responsible for roughly one-third of the events and is also present in the
theoretical distribution. This agreement with the theoretical distribution
strongly suggests that these features do not originate from a background of
$\alpha$ + \textit{p} + \textit{p} events that are not associated with $^6$Be
decay. Even at such a high excitation energy, the decay is therefore not purely
sequential, and the contributions of the different decay mechanisms cannot be
completely disentangled. The democracy of the decay is preserved in the
sense that different parts of the kinematical space have comparable populations.

\textit{Discussion.}
%
%
--- The mechanism of three-body decays in nuclei is often discussed in terms
of either the ``diproton'' or ``sequential'' decay mechanisms. The present data
demonstrate two results which can be seen as paradoxical and reflect the
complexity of the problem:

\noindent (i) It is now well established that the pure diproton decay
mechanism is not a good picture for the $2p$ decay (\cite{Pfutzner:2012} and references
therein). However, this does not mean that the $p$-$p$ final-state
interaction is absolutely not important for the formation of correlation patterns in
such decays. In such a context then, diprotons are expected
to be important for the
lowest energies. However, in $^{6}$Be decay there is a very clear indication
that the formation of the low-energy $p$-$p$ correlation is enhanced as the
decay energy increases.
It is also more pronounced in the excited $2^+$ state compared to the ground
$0^+$ state.

\noindent  (ii) In $^{6}$Be, the accessibility of the broad intermediate states
in the energy window of the three-body decay first leads to what appears as a
\emph{suppression of the sequential decay mechanism} in favor of three-body
democratic dynamics. Only at decay energies $E_T \gtrsim 2 E_r$($^5$Li)+$\Gamma
$($^5$Li) do the signs of sequential decay become visible in the correlation
patterns. However, even at such energies, the actual mechanism is a complex
mixture of contributions of $\alpha$-$p$ and $p$-$p$ final-state interactions, which
cannot be disentangled. Some indications for this decay complexity were found in Ref.\ 
\cite{Pfutzner:2012} based on simplified theoretical models. Now we have a
strong confirmation of this finding. This establishes the validity of democratic
decay as an appropriate description of the decay mechanism in a much broader
energy range than ever expected.


\textit{Conclusions.}
%
%
--- High-statistical-significance and high-resolution three-body correlation data were
obtained for $^{6}$Be decay over a broad range of decay energies. These
experimental results are reproduced by the three-cluster  model. The data
elucidate the mechanism of democratic decay and emphasize the paradoxical
and rather complex nature of three-body decay. They completely devalue the
simplistic ideas of sequential and diproton decay in favor of complex
three-body dynamics.

%
\textit{Acknowledgments.}
%
%
--- I.A.E.\ and L.V.G.\ are supported by the Helmholtz Association under Grant
Agreement IK-RU-002 via the FAIR-Russia Research Center. L.V.G.\ is supported by HIC for FAIR research grant, Russian RFBR Grant No.\ 11-02-00657-a, and Russian Ministry of Education and Science Grant No.\ NS-215.2012.2. I.A.E. was supported in part by German BMBF via the Heisenberg-Landau program. This work
was supported by the U.S. Department of Energy, Division of Nuclear Physics, 
under Grants No. DE-FG02-87ER-40316 and No. DE-FG02-04ER41320 and the National
Science Foundation under Grants No. PHY-0606007 and No. PHY-9977707.




\begin{thebibliography}{99}



\bibitem{Bochkarev:1989}
O.V.~Bochkarev \textit{et al.},
Nucl.\ Phys.\ \textbf{A505} 215, (1989).


\bibitem{Pfutzner:2012}
M.~Pf\"utzner, L.V.~Grigorenko, M.~Karny, and K.~Riisager,
Rev.\ Mod.\ Phys.\ \textbf{84}, 567 (2012).

\bibitem{Danilin:1987}
B.V.~Danilin, M.V.~Zhukov, A.A.~Korsheninnikov, L.V.~Chulkov,
and V.D.~Efros,
Yad.\ Fiz.\ \textbf{46},  427 (1987) 
[Sov.\ J.\ Nucl.\ Phys.\ \textbf{46}, 225 (1987)].

\bibitem{Barker:2003}
F.C.~Barker,
Phys.\ Rev.\ C \textbf{68}, 054602 (2003).


\bibitem{Blank:2008}
B.~Blank, and M.~Ploszajczak,
Rep.\ Prog.\ Phys.\ \textbf{71}, 046301 (2008).


\bibitem{Alvarez:2008}
R.~Alvarez-Rodriguez, H.O.U.~Fynbo, A.S.~Jensen,and  E.~Garrido,
Phys.\ Rev.\ Lett.\ \textbf{100}, 192501 (2008).


\bibitem{Grigorenko:2009}
L.V.~Grigorenko \textit{et al.},
Phys.\ Lett.\ B \textbf{677}, 30 (2009).


\bibitem{Grigorenko:2009c}
L.V.~Grigorenko \textit{et al.},
Phys.\ Rev.\ C \textbf{80}, 034602 (2009).


\bibitem{Charity:2010}
R.J.~Charity \textit{et al.},
Phys.\ Rev.\ C \textbf{82}, 041304 (2010).


\bibitem{Papka:2010}
P. Papka \textit{et al.},
Phys.\ Rev.\ C \textbf{81}, 054308 (2010).


\bibitem{Fomichev:2012}
A.S.~Fomichev \textit{et al.},
Phys.\ Lett.\ B \textbf{708} 6, (2012).


\bibitem{Wallace:2007}
M.S.~Wallace \textit{et al.},
Nucl.\ Instrum.\ Methods Phys.\ Res., Sect.\ A \textbf{583}, 302 (2007).


\bibitem{Engel:2007}
G.L.~Engel \textit{et al.},
Nucl.\ Instrum.\ Methods Phys.\ Res., Sect.\ A \textbf{573}, 418 (2007).


\bibitem{Zhukov:1994}
M.V.~Zhukov, A.A.~Korsheninnikov, and M.H.~Smedberg,
Phys.\ Rev.\ C \textbf{50}, R1 (1994).


\bibitem{Geesaman:1977}
D.F.~Geesaman  \textit{et al.},
Phys.\ Rev.\ C \textbf{15}, 1835 (1977).






\end{thebibliography}
\end{document}